\documentclass[12pt]{article}
\usepackage{a4,amsmath}
\renewcommand{\baselinestretch}{1.1}
\newcommand{\PP}{\mathcal{P}}
\newcommand{\TT}{\mathcal{T}}
\newcommand{\CC}{\mathcal{C}}
\begin{document}
\begin{center}
%
{\Huge Comment on}\\\vspace{0.5\baselineskip}
{\Huge `Must a Hamiltonian be hermitian'}

\vspace{2\baselineskip}
{\Large Andr\'e van Hameren%
}

\vspace{0.25\baselineskip}
{\it\large Institute of Nuclear Physics, NCSR Demokritos, 15310 Athens, Greece}

\vspace{0.25\baselineskip}
{\tt\large andrevh@inp.demokritos.gr}

\vspace{0.25\baselineskip}
{\large\today}

\renewcommand{\baselinestretch}{1}
\vspace{2\baselineskip}
{\bf Abstract}\\\vspace{0.5\baselineskip}
\parbox{0.8\linewidth}{\small\hspace{15pt}%
A small comment on the paper with the mentioned title by Carl M.\ Bender, Dorje C.\ Brody and Hugh F.\ Jones.
}
\end{center}
\vspace{\baselineskip}

As argued in \cite{paper}, the eigenfunctions $\phi_n$ of the Sturm-Liouville
eigenvalue problem $(8)$ are fixed up to a constant phase factor.  
Instead of $\phi_n$, one can also choose
\begin{displaymath}
   \psi_n(x) = i^n\phi_n(x)
\;\;,
\end{displaymath}
to be the eigenfunctions under considerations, which satisfy 
\begin{displaymath}
   \PP\TT\psi_n(x) = \psi_n^*(-x) = (-i)^n\phi_n^*(-x) 
                   = (-i)^n\phi_n(x) = (-)^n\psi_n(x)
\;\;,
\end{displaymath}
and 
\begin{displaymath}
   \delta(x-y) = \sum_n(-)^n\phi_n(x)\phi_n(y) = \sum_n\psi_n(x)\psi_n(y)
\;\;.
\end{displaymath}
This formula, then, looks more familiar than $(6)$. Only if the functions
$\psi_n$ are non-real, one usually has a different formula:
$\sum_n\psi_n^*(x)\psi_n(y) = \delta(x-y)$.
The operator $\CC$ can be represented by
\begin{displaymath}
   \CC(x,y) = \sum_n(-)^n\psi_n(x)\psi_n(y)
\;\;.
\end{displaymath}
It acts on the functions $\psi_n$ as $\CC\psi_n=(-)^n\psi_n$, so that they
are invariant under $\CC\PP\TT$:
\begin{displaymath}
   \CC\PP\TT\psi_n = \psi_n
\;\;,
\end{displaymath}
and are orthonormal under
\begin{displaymath}
   \langle \psi_n|\psi_m \rangle
   = \int[\CC\PP\TT\psi_n(x)]\,\psi_m(x)\,dx
   = \int\psi_n(x)\,\psi_m(x)\,dx
\;\;,
\end{displaymath}
which looks like a {\em real\/} inner product%
\footnote{{\it i.e.\/} $\langle f|g\rangle=\langle g|f\rangle$ instead of 
$\langle f|g\rangle=\langle g|f\rangle^*$}. 
So we see that $\CC\PP\TT$-invariance of a Hamiltonian, as introduced in \cite{paper},
is equivalent with the
existence of a set of eigenfunctions that are complete and orthonormal under a
{\em real\/} inner product.  The (complex) inner product in the whole Hilbert
space can be defined by
\begin{displaymath}
   \langle f|g\rangle = \sum_nf_n^*g_n
   \quad\textrm{with}\quad
   f_n = \int\psi_n(x)\,f(x)\,dx
   \;\;,\;\;
   g_n = \int\psi_n(x)\,g(x)\,dx
\;\;.
\end{displaymath}

 
%

%
\end{document}